\documentclass[twocolumn]{aastex631}

\usepackage{amsfonts}
\usepackage{amsmath}
\usepackage{amssymb}
\usepackage{subfigure}
\usepackage{graphicx}
\usepackage{epstopdf}
\usepackage{epsfig}
\usepackage{float}
\usepackage{hyperref}
\usepackage{color}
\usepackage{ulem}
\usepackage{cancel}
\usepackage{mathrsfs}
\usepackage{booktabs}
\usepackage{multirow}
\usepackage{natbib}
\usepackage{ulem}

\begin{document}

\title{Scattering Cross Sections of Magnetized Particles within Intense Electromagnetic Waves: Application to Fast Radio Bursts}

\author[0009-0009-7749-8998]{Yu-Chen Huang}
\affiliation{Department of Astronomy, University of Science and Technology of China, Hefei 230026, China; daizg@ustc.edu.cn}
\affiliation{School of Astronomy and Space Science, University of Science and Technology of China, Hefei 230026, China}

\author[0000-0002-1766-6947]{Shu-Qing Zhong}
\affiliation{Department of Astronomy, University of Science and Technology of China, Hefei 230026, China; daizg@ustc.edu.cn}
\affiliation{School of Astronomy and Space Science, University of Science and Technology of China, Hefei 230026, China}

\author[0000-0002-7835-8585]{Zi-Gao Dai}
\affiliation{Department of Astronomy, University of Science and Technology of China, Hefei 230026, China; daizg@ustc.edu.cn}
\affiliation{School of Astronomy and Space Science, University of Science and Technology of China, Hefei 230026, China}

\begin{abstract}

	Recently, Beloborodov suggested that there exists a resonance phenomenon between an extremely intense electromagnetic wave and internal magnetized particles. The particles exchange energy with the wave at frequent resonance events and then reach the radiation reaction limit immediately. This process greatly enhances the scattering cross section of the particles. Note that these results only involve an extraordinary (X) mode wave. In this paper, we focus on an intense ordinary (O) mode wave propagating through magnetized particles and compare it with the case of the X-mode wave. Our result shows that the scattering cross section of the particles in the O-mode wave is significantly smaller than that in the X-mode wave. This has important implications for the transparency of a fast radio burst (FRB) inside the magnetosphere of a magnetar. We argue that there is a strong scattering region in the stellar magnetosphere, within which an O-mode wave is more transparent than an X-mode wave for an FRB.

\end{abstract}




\section{Introduction}

Some highly magnetized neutron stars, known as magnetars, can generate extremely intense transient radio emissions called fast radio bursts (FRBs) \citep{Lorimer2007,Bochenek2020,CHIME/FRBCollaboration2020}. Currently, the radiation mechanism of FRBs is not well understood. Even in the magnetar models, there are  still different viewpoints on the location where FRBs are generated. Some models propose that FRBs could be produced inside the magnetosphere \citep{Yang2018,Lu2020,Wang2022,Zhang2022,Liu2023}, while some other models suggest that FRBs could take place outside the magnetosphere \citep{Lyubarsky2014,Beloborodov2017,Beloborodov2020,Metzger2019,Margalit2020}. It was also proposed that FRBs could be generated near the light cylinder \citep{Lyubarsky2020,Mahlmann2022}. More details about the models can be seen in some recent reviews \citep{Platts2019,Lyubarsky2021,Xiao2021,Zhang2023a}.

The typical luminosity of cosmological FRBs is approximately 10 orders of magnitude larger than that of regular radio pulses from pulsars. Therefore, a strong wave effect in FRBs likely introduces new characteristics distinct from those radio pulses and can constrain theoretical models \citep{Kumar2020,Yang2020}. To describe this effect, a nonlinear parameter can be introduced, which is defined as
\begin{equation}
	a_0=\frac{eE_0}{mc\omega}\approx2.3\times10^4 \nu_{9}^{-1} L_{\text{iso},42}^{1/2}r_{9}^{-1},
	\label{a0}
\end{equation}
where $c$ is the speed of light, and $e$ and $m$ are electron charge and mass. $E_0$, $L_{\text{iso}}=(10^{42}\text{ erg s}^{-1})L_{\text{iso},42}$, and $\nu=\omega/2\pi=(10^9\text{ Hz})\nu_{9}$  represent the electric field strength, isotropic luminosity, and frequency of an FRB, respectively. $r=(10^9\text{ cm})r_9$ is the magnitude of the light cylinder radius,
\begin{equation}
	R_{\text{LC}}=\frac{cP}{2\pi}\approx\left(4.8\times 10^9\text{ cm}\right)\left(\frac{P}{1\text{ s}}\right),
\end{equation}
where $P$ is the period of the magnetar. The nonlinear parameter in Eq. (\ref{a0}) can also be written as $a_0=v_{\text{ec}}/c$, where $v_{\text{ec}}=eE_0/m\omega$ can be understood as the nonrelativistic characteristic oscillation velocity of an electron in the wave. For a strong wave with $a_0\gg 1$, the classical nonrelativistic theorem describing the Thomson scattering fails, and the strong wave effect is supposed to be taken into account. The strong wave effect of FRBs has been extensively discussed in the near-source environment \citep{Luan2014,Gruzinov2019a,Lu2020a,Yang2020}, while there have been few works focusing on the effect within the magnetosphere.

Recently, \cite{Beloborodov2022} proposed a novel description of the scattering of magnetized particles in such a strong electromagnetic wave. For instance, an FRB traveling outward through the magnetosphere of a magnetar likely encounters a region with
\begin{equation}
	1<\frac{\omega_B}{\omega}<a_0,
	\label{ssregion}
\end{equation}
where $\omega_B=eB_{\text{bg}}/mc$ is the electron cyclotron frequency in a background magnetic field. This condition can be satisfied near the outer magnetosphere where
\begin{equation}
	\frac{\omega_B}{\omega}\approx2.8\times10^3 B_{\text{s},15}\nu_{9}^{-1}r_{9}^{-3}.
	\label{omegabomega}
\end{equation}
We adopt a dipole background field with surface strength $B_{\text{s}}=(10^{15}\text{ G})B_{\text{s},15}$. Within the region where Eq. (\ref{ssregion}) holds, the wave exchanges energy with internal charged particles at resonance events, and pushes the particles to the radiation reaction limit. In this case, the energy loss of the particles due to radiation equals the energy gained from resonance. The scattering cross section of particles can then be semianalytically derived and is found to be significantly enhanced compared with the classical Thomson scattering.

We note that the above consideration only involves an extraordinary (X) mode wave, whose electric field is perpendicular to the propagation direction and the background magnetic field. In fact, both ordinary (O) mode and X-mode waves can be generated within the framework of magnetospheric coherent emission models, such as curvature radiation \citep{Wang2012,Kumar2017,Yang2020a}. It is therefore necessary to investigate the scattering of magnetized particles in an intense O-mode wave, whose electric field lies in the plane defined by the propagation direction and the background magnetic field.

In this paper, we first solve the equation of motion of a charged particle in an O-mode wave and an X-mode wave, respectively. We then study the resonance phenomenon between the wave and the particle in both modes in Section \ref{pmwm}. The derivation of the cross sections in different modes is conducted in Section \ref{scsdp}. Based on these results, we further investigate the implications for FRBs in Section \ref{ifrb}. The final summary and discussion are presented in Section \ref{discussion}. In this paper, the notation $Q_n=Q/10^n$ and cgs units are adopted.

\section{Particle Motion in an Intense Wave within a Background Magnetic Field}\label{pmwm}

\begin{figure}
	\begin{center}
		\subfigure{
			\includegraphics[width=0.38\textwidth]{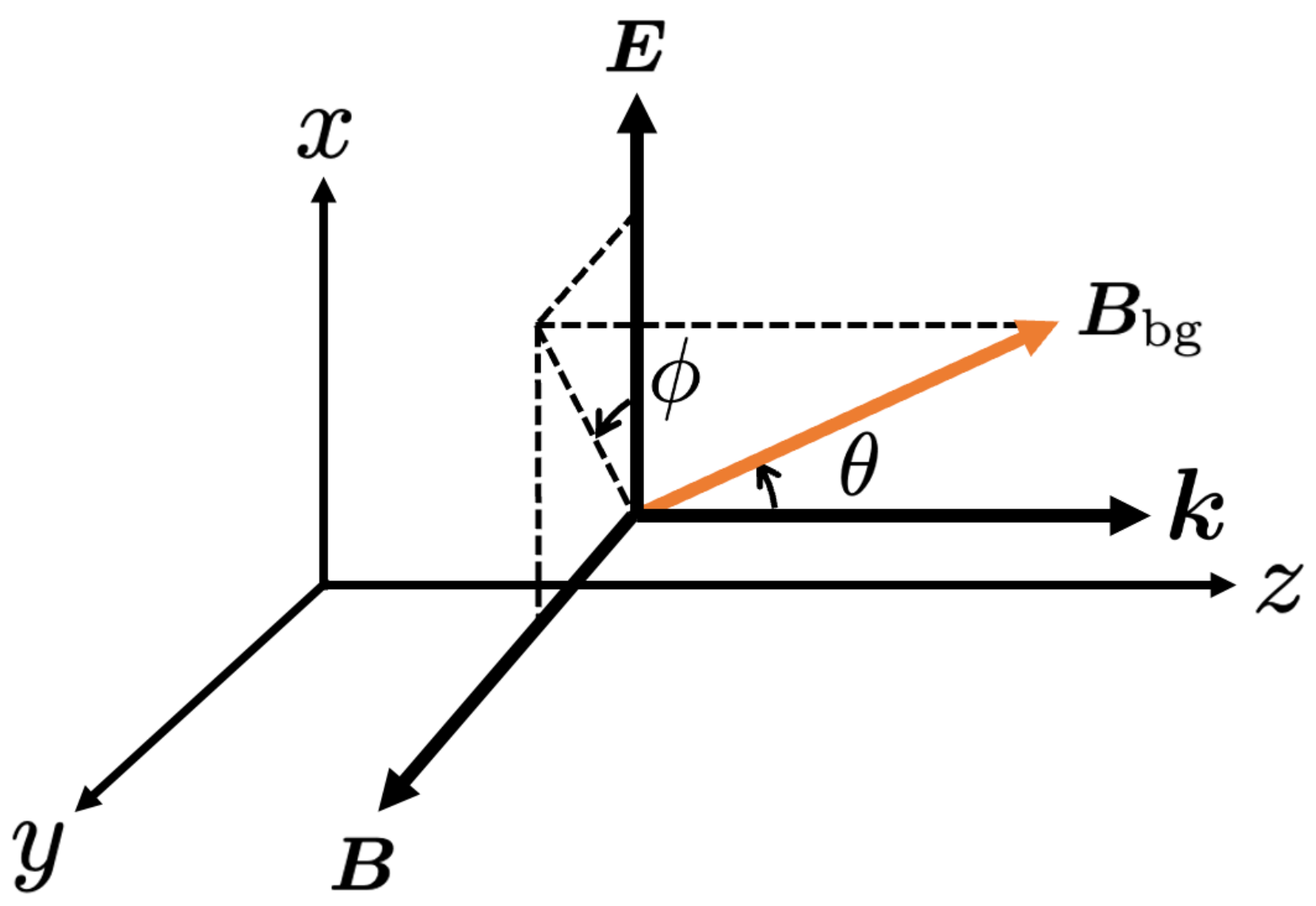}}
		\caption{Schematic illustration of an electromagnetic wave propagating along the $z$-axis. The electric field and magnetic field of the wave are along the $x$- and $y$-axis, respectively. The direction of the background magnetic field is determined by parameters $\theta$ and $\phi$.}	
		\label{embg}
	\end{center}
\end{figure}

The equation of motion of a charged particle within an intense electromagnetic wave and a background magnetic field is
\begin{equation}
	\frac{d\boldsymbol{u}}{dt}=\frac{e}{mc}\left(\boldsymbol{E}+\boldsymbol{\beta}\times\boldsymbol{B}+\boldsymbol{\beta}\times\boldsymbol{B}_{\text{bg}}\right),
	\label{emparticle}
\end{equation}
where $\boldsymbol{u}=\gamma\boldsymbol{\beta}$ is the spatial component of the normalized four-velocity $u^{\alpha}=dx^{\alpha}/cd\tau=(\gamma,\gamma\boldsymbol{\beta})$. We assume that the incident wave is linearly polarized. As shown in Figure \ref{embg}, its electric field, magnetic field, and propagation direction are along the $x$-, $y$- and $z$-axes, respectively. For simplicity, we also assume that the incident wave is a monochromatic plane wave, i.e., $e\boldsymbol{E}/mc\omega=\boldsymbol{e_x}a_0\sin{(\omega\xi)}$ and $e\boldsymbol{B}/mc\omega=\boldsymbol{e_y}a_0 \sin{(\omega\xi)}$, where $\xi\equiv t-z/c$. The background magnetic field can point in any direction, which is given by $\boldsymbol{B}_{\text{bg}}=B_{\text{bg}}(\sin{\theta}\cos{\phi}, \sin{\theta}\sin{\phi},\cos{\theta})$.

\begin{figure*}
	\begin{center}
		\subfigure{
			\includegraphics[width=0.32\textwidth]{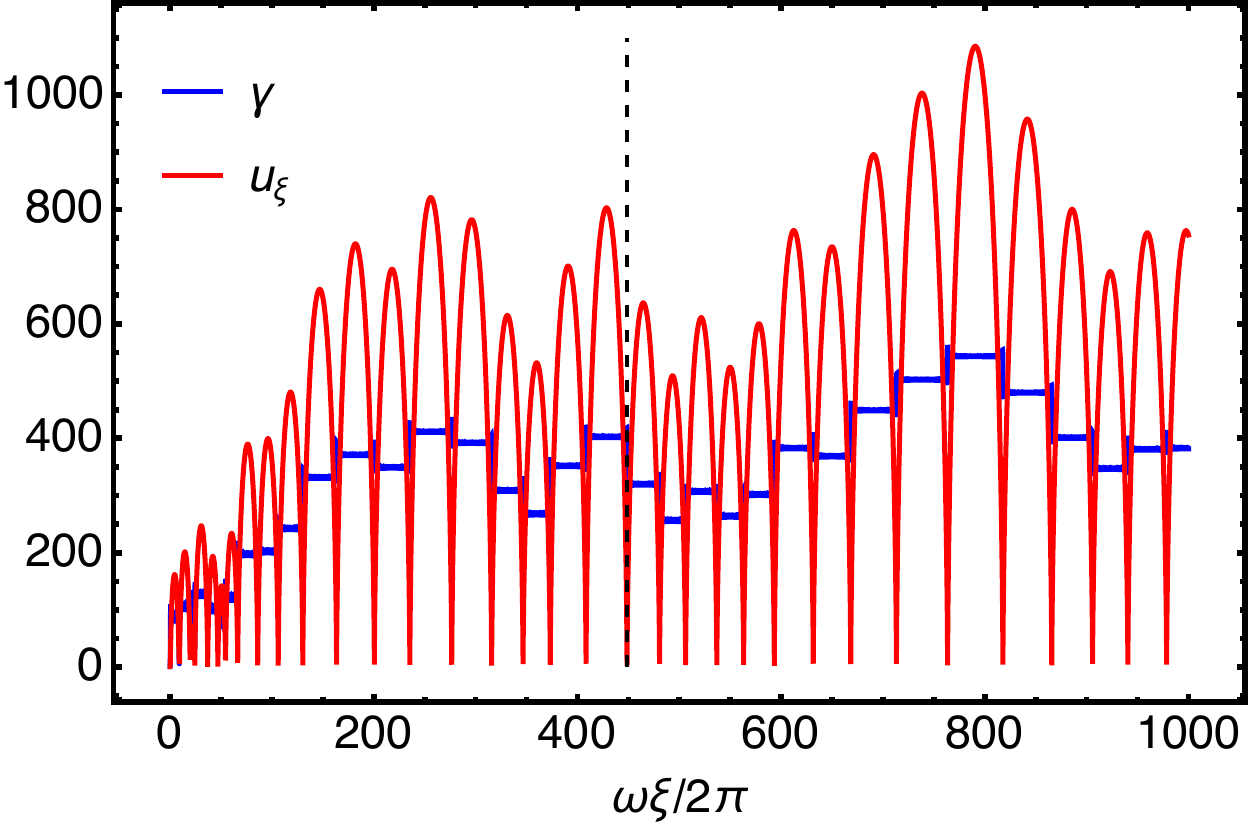}}
		\subfigure{
			\includegraphics[width=0.32\textwidth]{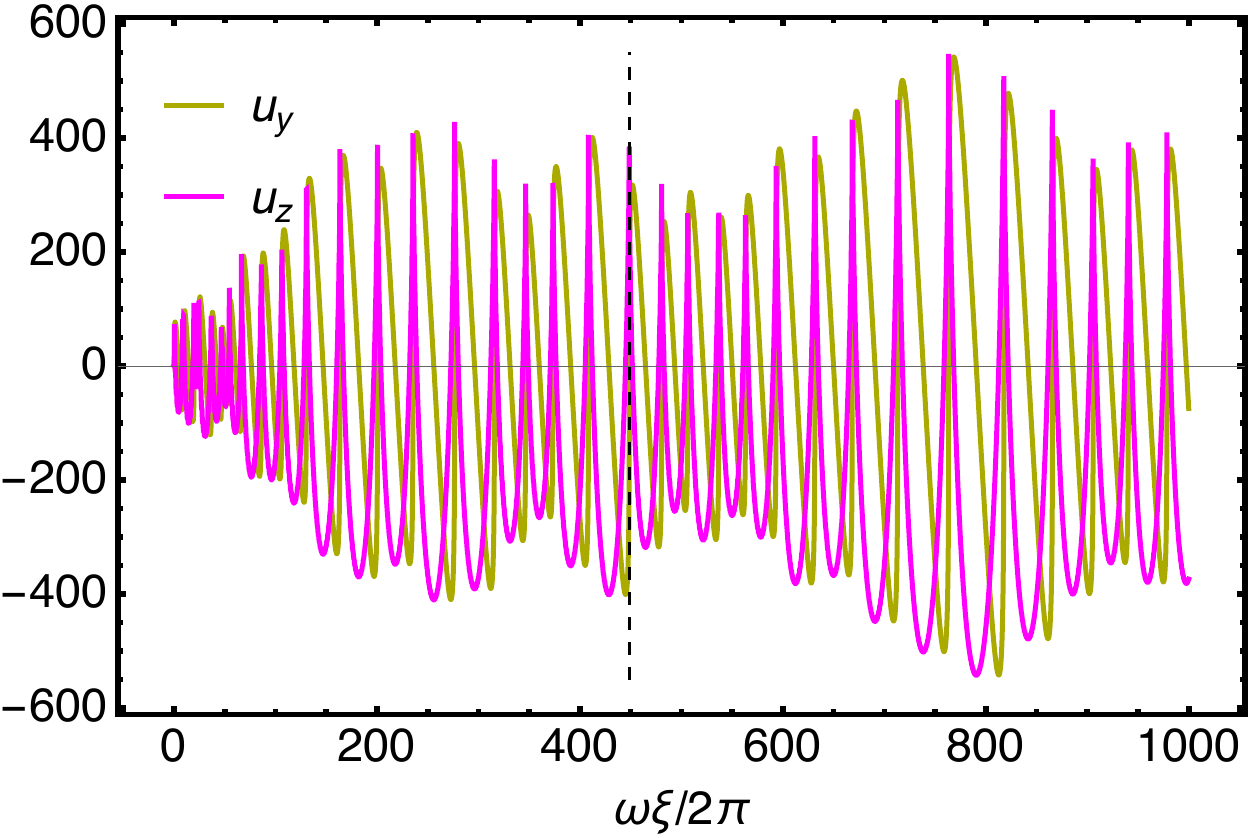}}
		\subfigure{
			\includegraphics[width=0.32\textwidth]{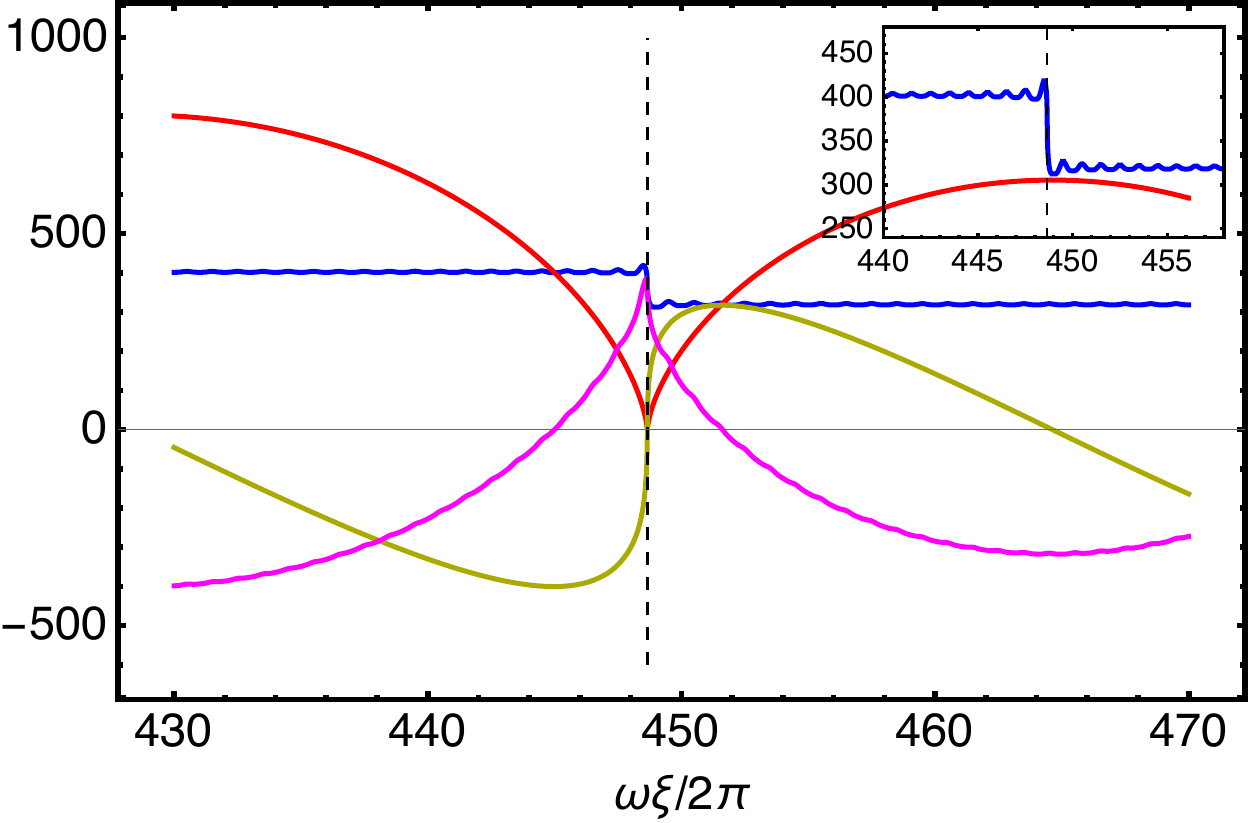}}
		\subfigure{
			\includegraphics[width=0.32\textwidth]{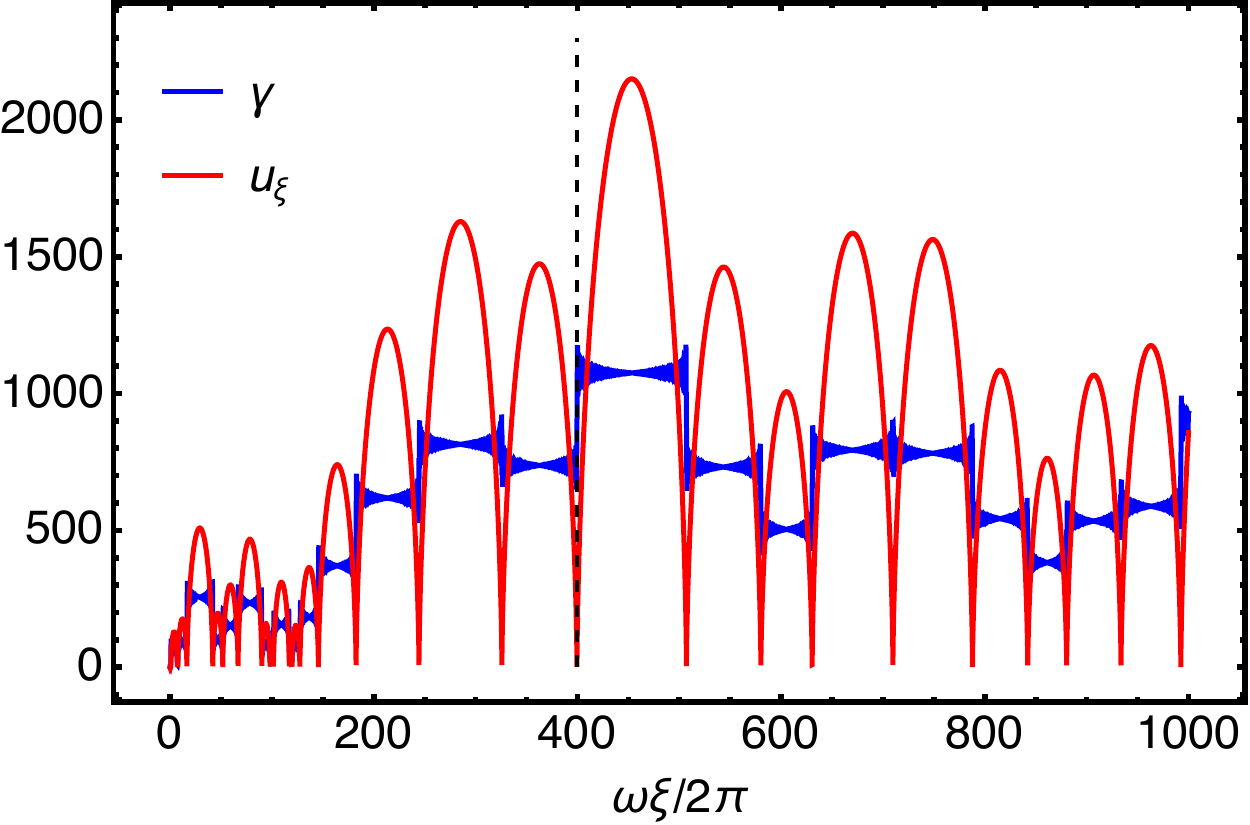}}
		\subfigure{
			\includegraphics[width=0.32\textwidth]{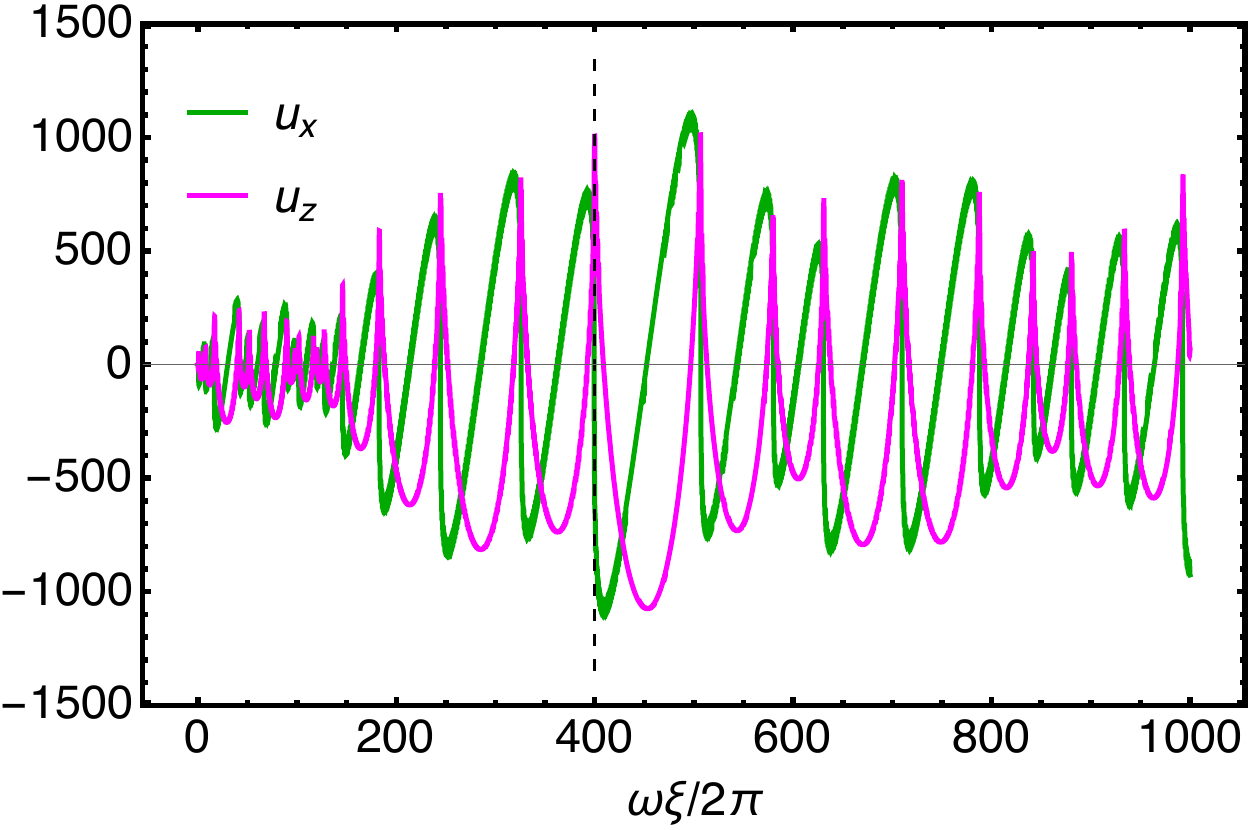}}
		\subfigure{
			\includegraphics[width=0.32\textwidth]{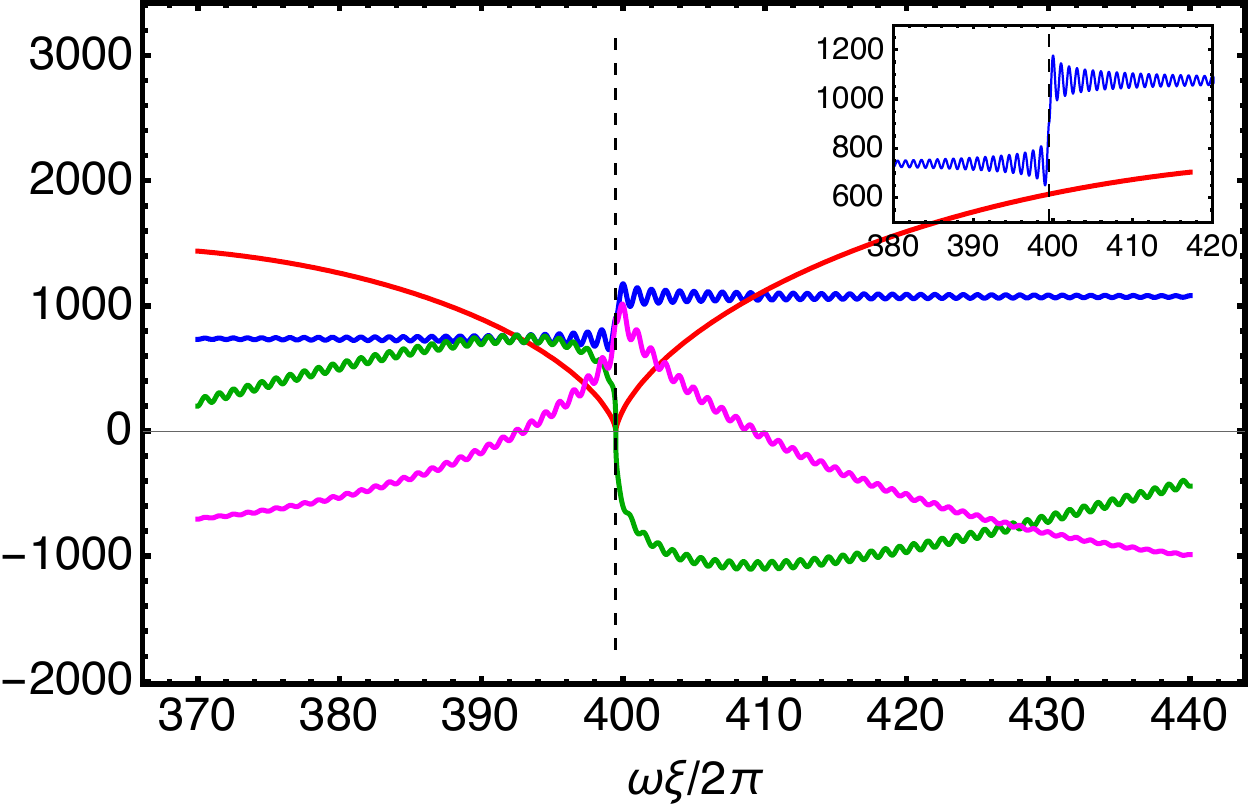}}
		\caption{Solutions of motion of a charged particle in an intense electromagnetic wave and a background magnetic field. The parameters $a_0=30$ and $\omega_B/\omega=10$ are adopted. In the upper panels, the incident wave is O mode. The angle parameters are $\theta=\pi/2$ and $\phi=0$. In the lower panels, the wave is X mode. The angle parameters are $\theta=\pi/2$ and $\phi=\pi/2$. The vertical black dashed lines refer to the position of a resonance event. The left two panels exhibit the pumping effect of the Lorentz factor during several large-amplitude oscillations. The middle two panels exhibit the components of the normalized four-velocity. The right two panels exhibit the behavior of the particle near a resonance event in detail, in which the insets show the pumping effect of the Lorentz factor.}
		\label{emsolution}
	\end{center}
\end{figure*}

Eq.  (\ref{emparticle}) can be written as three scalar equations:
\begin{equation}
	\begin{aligned}
		\frac{dU_x}{dt}&=\frac{\omega_B}{\gamma}\left(u_y  \cos{\theta}-u_z \sin{\theta}\sin{\phi}\right),\\
		\frac{du_y}{dt}&= \frac{\omega_B}{\gamma}\left(u_z  \sin{\theta}\cos{\phi}-u_x \cos{\theta}\right),\\
		\frac{du_\xi}{dt}&=\frac{\omega_B}{\gamma}\left(u_y \sin{\theta}\cos{\phi}-u_x \sin{\theta}\sin{\phi}\right),
	\end{aligned}\label{threeequamotion}
\end{equation}
where $U_x=u_x+a_0\cos{(\omega\xi)}$ and $u_\xi=d\xi/d\tau=\gamma-u_z$. The solutions of motion of an initial rest particle are exhibited in Figure \ref{emsolution}, in which the wave is O-mode (upper panels) or X-mode (lower panels). The motion of the particle can be regarded as the superposition of a large-amplitude oscillation and a small-amplitude oscillation, as shown in the right two panels in Figure \ref{emsolution}. According to Eq. (\ref{threeequamotion}), we can infer and estimate that the frequency of the large-amplitude oscillation is the Larmor frequency $\omega_\text{L}=\omega_B/\gamma$. The small-amplitude oscillation is driven directly by the electromagnetic field of the wave. The wave undergoes one cycle with $\omega\delta\xi=\omega\left(1-\beta_z\right)\delta t\sim2\pi$, providing an approximate angular frequency $\omega\left(1-\beta_z\right)$ for the small-amplitude oscillation. In the O-mode wave, the motion of the particle in the $x$-axis direction can be solved analytically, i.e., $u_x=a_0\left[1-\cos{(\omega\xi)}\right]$. In the X-mode wave, the motion of the particle in the $y$-axis direction vanishes, i.e., $u_y=0$.

Figure \ref{emsolution} also shows that the energy of the particle can be pumped to a different value at every resonance event. This pumping effect was found by \cite{Beloborodov2022}. We find that the value of the pumped Lorentz factor of the particle in an O-mode wave is apparently different from that in an X-mode wave. As we demonstrate below, the gained Lorentz factors of the particle at resonance in these two modes are totally different.

\begin{figure*}
	\begin{center}
		\subfigure{
			\includegraphics[width=0.38\textwidth]{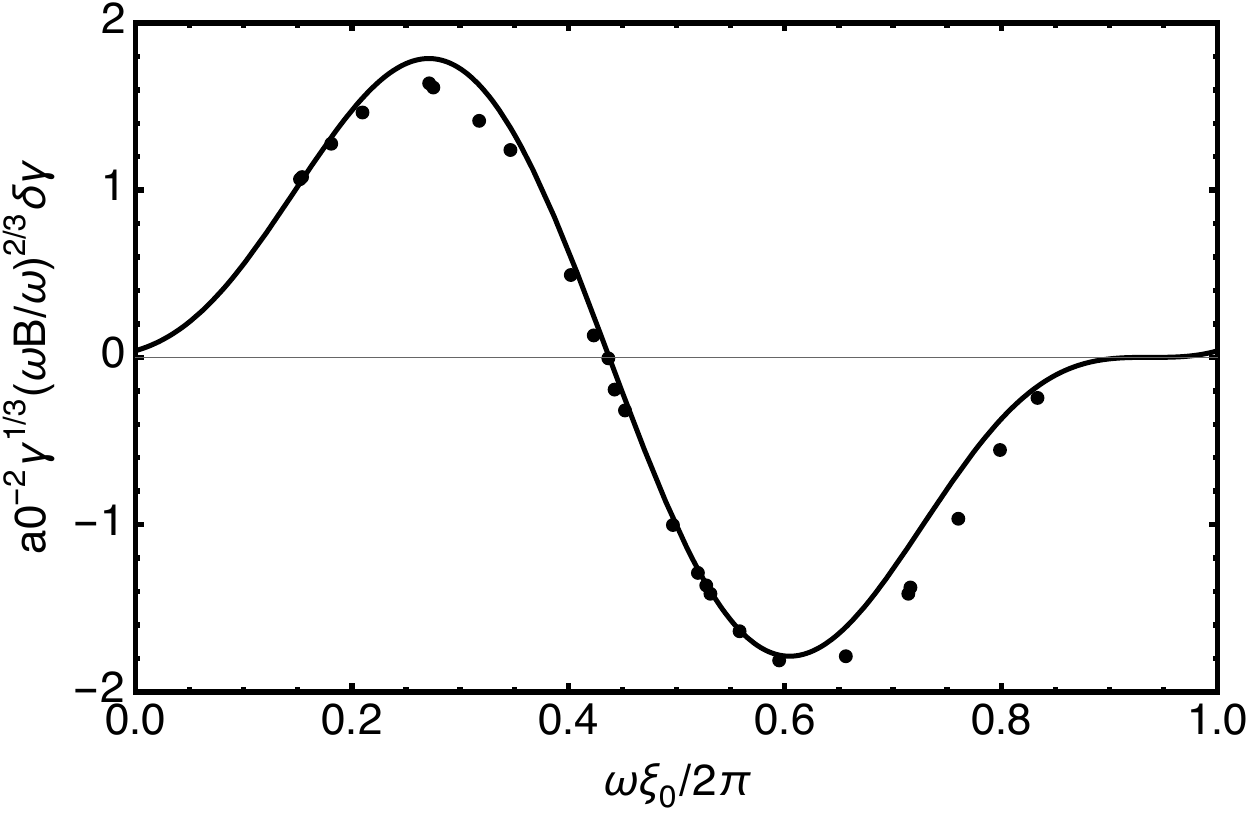}}
		\subfigure{
			\includegraphics[width=0.38\textwidth]{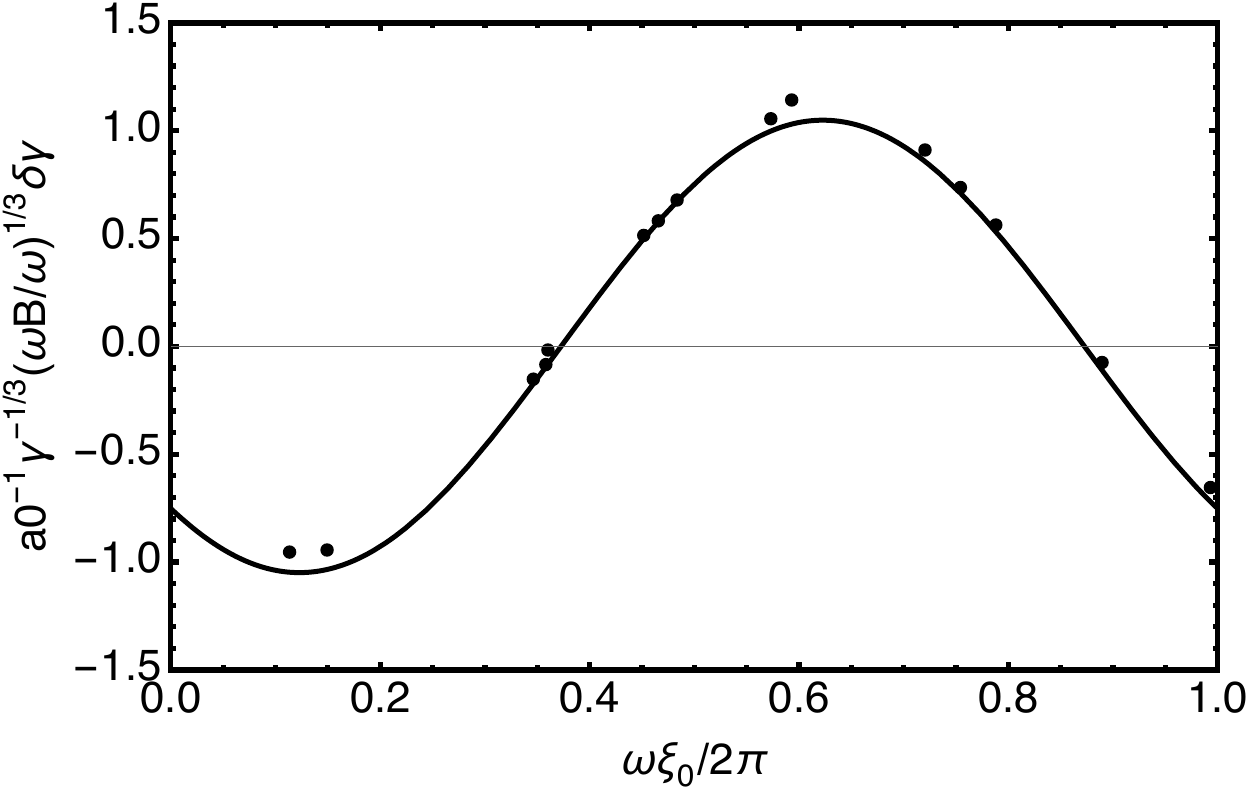}}
		\caption{The rescaled pumped Lorentz factor of the particle vs. the phase of the wave at resonance. In the O-mode wave, $a_0^{-2}\gamma^{1/3}(\omega_B/\omega)^{2/3}\delta\gamma\approx H_O  \sin{(\omega\xi_0)}\left[1-\cos{(\omega\xi_0)}\right]$ with $H_O\approx1.39$ (left panel). In the X-mode wave, $a_0^{-1}\gamma^{-1/3}(\omega_B/\omega)^{1/3}\delta\gamma\approx H_X\sin{(\omega\xi_0)}$ with $H_X\approx1.05$ (right panel).}	
		\label{deltagamma}
	\end{center}
\end{figure*}

The pumped energy of a particle at a resonance event is provided by the electric field of the wave, that is, $mc d\gamma/dt=eE \beta_x$.
This equation can be equivalently written as
\begin{equation}
	d\gamma=\frac{a_0}{\gamma}\frac{u_x \sin{(\omega\xi)} }{1-\beta_z}\omega d\xi.
	\label{dgammageneral}
\end{equation}
In the vicinity of resonance, $\beta_{z}^{\text{res}}\sim\cos{\delta\psi}\sim1-\delta\psi^2/2$, where $\delta\psi=\omega_{\text{L}}\delta t$. The resonance event occurs within one small-amplitude oscillation, i.e., $\omega(1-\beta_z)\delta t\sim 2\pi $. Therefore, one arrives at
\begin{equation}
	\delta\psi\sim\left(\frac{\omega_{\text{L}}}{\omega}\right)^{1/3}.
\end{equation}
For the particle in the O-mode wave, one has $u_x=a_0\left[1-\cos{(\omega\xi)}\right]$. Substituting these relations into Eq. (\ref{dgammageneral}) gives the pumped Lorentz factor at a resonance event
\begin{equation}
	\delta\gamma_O\approx H_O a_0^2 \gamma^{-1/3}\left(\frac{\omega_B}{\omega}\right)^{-2/3}\sin{(\omega\xi_0)}\left[1-\cos{(\omega\xi_0)}\right],
	\label{dgammaHO}
\end{equation}
where $\omega\xi_0$ is the phase of the wave at resonance. The coefficient $H_O\approx1.39$, which can be obtained by fitting the data of $a_0^{-2}\gamma^{1/3}(\omega_B/\omega)^{2/3}\delta\gamma$ and $\sin{(\omega\xi_0)}\left[1-\cos{(\omega\xi_0)}\right]$ at resonance events in the numerical solution, as shown in the left panel of Figure \ref{deltagamma}. For the particle in the X-mode wave, $u_x$ near a resonance event can be estimated as 
\begin{equation}
	\begin{aligned}
		u_x^{\text{res}}&=\left[\gamma^2 (1-\beta_z^{2})-1\right]^{1/2}\sim\left(\gamma^2 \delta\psi^2-1\right)^{1/2}\\
		&\sim\left[\gamma^2\left(\frac{\omega_{\text{L}}}{\omega}\right)^{2/3}-1\right]^{1/2}\sim\gamma^{2/3}\left(\frac{\omega_B}{\omega}\right)^{1/3},
	\end{aligned}
\end{equation}
where we have used the relation $\gamma^2-u_x^2-u_z^2=1$. A further calculation of Eq. (\ref{dgammageneral}) gives
\begin{equation}
	\delta\gamma_X \approx H_X a_0 \gamma^{1/3}\left(\frac{\omega_B}{\omega}\right)^{-1/3}\sin{(\omega\xi_0)}.
	\label{dgammaHx}
\end{equation}
The fitting result is shown in the right panel of Figure \ref{deltagamma}, in which $H_X\approx 1.05$ is obtained. The derivation of Eq. (\ref{dgammageneral}) is independent of whether the wave is O mode or X mode. Therefore, one can observe that the difference in the pumped Lorentz factor between both modes is solely determined by $u_x$, i.e., the motion of the charged particle along the electric field direction. This motion behaves differently in O-mode and X-mode waves, resulting in a difference in the pumped Lorentz factor during resonance.

In a sufficiently long time, the pumping effect of the Lorentz factor behaves like a random walk process, and hence can theoretically boost the energy of the particle to high values. Nevertheless, the maximal Lorentz factor can be limited by radiation losses. A radiation reaction limit can be reached when the energy loss due to radiation equals to the energy gained  from resonance. In the following section, we demonstrate that after reaching such an equilibrium state, a significant difference in the Lorentz factors of the particle emerges between the O-mode and X-mode waves. The difference results in distinct scattering cross sections of the particle.

\section{Scattering Cross Sections in Different Polarized Modes}\label{scsdp}

The total emission power of a relativistic electron is \citep{Rybicki1991}
\begin{equation}
	P_{e}=\frac{2e^2}{3c^3}\gamma^4\left(a_{\perp}^2+\gamma^2 a_{\parallel}^2\right),
\end{equation}
where $a_{\perp}$ and $a_{\parallel}$ represent the acceleration perpendicular and parallel to the velocity, respectively. The emission power can be equivalently expressed in terms of the electromagnetic field as
\begin{equation}
	\dot{\gamma}_{\text{em}}mc^2=\frac{c}{4\pi}\sigma_T\left[ \left(\gamma\boldsymbol{E}+\boldsymbol{u}\times\boldsymbol{B}+\boldsymbol{u}\times\boldsymbol{B}_{\text{bg}}\right)^2-\left(\boldsymbol{u}\cdot\boldsymbol{E}\right)^2\right],
\end{equation}
where $\dot{\gamma}_{\text{em}}$ is the loss rate of Lorentz factor due to radiation. $\sigma_T=8\pi r_e^2/3$ is the Thomson scattering cross section, and $r_e=e^2/mc^2$ is the classical electron radius. When $B\gg B_{\text{bg}}$, one has $\dot{\gamma}_{\text{em}}\sim \sigma_T E^2 u_\xi^2  /4\pi m c^2$. Averaging this loss rate over the period of a small-amplitude oscillation gives
\begin{equation}
	\overline{\dot{\gamma}}_{\text{em}}\sim\frac{1}{3}\frac{r_e a_0^2 \omega^2 }{c}u_\xi^2,
\end{equation}
where $u_\xi^2\approx \overline{u_\xi^2}$ has been used.

The radiation reaction limit requires that, within each large-amplitude oscillation, the energy loss due to radiation equals to the energy gained from resonance, i.e., $\langle\dot{\gamma}_{\text{em}}\rangle\Delta t=\delta\gamma$, where the angle bracket denotes the time average during one large-amplitude oscillation, with a timescale $\Delta t\sim 2\pi/\omega_{\text{L}}$. The average radiative loss rate $\langle\dot{\gamma}_{\text{em}}\rangle=r_e a_0^2 \omega^2 \langle u_\xi^2\rangle /3c\sim r_e a_0^2 \omega^2 \gamma^2 /3c$. Therefore, in the O-mode wave, the Lorentz factor at the radiation reaction limit is
\begin{equation}
	\gamma_{\text{rrl},O}\approx\left(\frac{3H_O}{2\pi}\frac{c}{r_e  \omega}\right)^{3/10}\left(\frac{\omega_B}{\omega}\right)^{1/10}.
	\label{gammarrlo}
\end{equation}
For the X-mode wave, the Lorentz factor at radiation reaction limit
\begin{equation}
	\gamma_{\text{rrl},X}\approx\left(\frac{3H_X}{2\pi}\frac{c}{r_e a_0 \omega}\right)^{3/8}\left(\frac{\omega_B}{\omega}\right)^{1/4}.
	\label{gammarrlx}
\end{equation}
Notice that the condition $\langle\dot{\gamma}_{\text{em}}\rangle\Delta t<\gamma$ must be satisfied to ensure that the energy loss during each large-amplitude oscillation is smaller than the total energy. This condition implies that there exists an upper limit for the particle Lorentz factor in both modes, i.e.,
\begin{equation}
	\gamma<\gamma_{\text{ul}}=\left(\frac{3}{2\pi}\frac{c}{r_e a_0^2 \omega}\right)^{1/2}\left(\frac{\omega_B}{\omega}\right)^{1/2}.
	\label{gammaulimit}
\end{equation}
Substituting Eqs. (\ref{gammarrlo}) and (\ref{gammarrlx}) into Eq. (\ref{gammaulimit}) gives $a_0<H_O^{-3/10}a_c\approx a_c$ for the O mode, and $a_0< H_X^{-3/5}a_c\approx a_c$ for the X mode, where a characteristic nonlinear parameter is defined as
\begin{equation}
	a_c=\left(\frac{3c\omega_B^2}{2\pi r_e \omega^3}\right)^{1/5}.
	\label{ac}
\end{equation}

The pumping effect of the Lorentz factor resembles a random walk process, and therefore the time to reach the radiation reaction limit can be simply obtained through an analogy with the random walk process, i.e., $t_{\text{rrl}}\sim\left(\gamma_{\text{rrl}}/\delta\gamma\right)^2 2\pi/\omega_{\text{L}}$, which gives
\begin{equation}
	\frac{\omega t_{\text{rrl},O}}{2\pi}\sim H_O^{-9/10}\left(\frac{a_0}{\omega_B/\omega}\right)^{3/2}\left(\frac{a_{c}}{a_0}\right)^{11/2}
\end{equation}
for the O-mode wave, and 
\begin{equation}
	\frac{\omega t_{\text{rrl},X}}{2\pi}\sim H_X^{-9/8}\left(\frac{a_0}{\omega_B/\omega}\right)^{3/2}\left(\frac{a_{c}}{a_0}\right)^{35/8}
\end{equation}
for the X-mode wave. Since the timescale is short, once particles are within the wave, they can reach the radiation reaction limit immediately.

The scattering cross section is defined as the time-averaged radiation power over the incident flux of the wave,
\begin{equation}
	\sigma=\frac{\langle\dot{\gamma}_{\text{em}}\rangle mc^2}{\langle S\rangle}\sim \gamma^2 \sigma_T,
\end{equation}
where we have used $\langle S\rangle=m c a_0^2 \omega^2/8\pi r_e$. Due to the short time to reach the radiation reaction limit, the scattering cross section of a particle in the O-mode wave can be estimated as
\begin{equation}
	\sigma_{O}\sim\gamma_{\text{rrl},O}^2 \sigma_T \approx\left(\frac{3H_O}{2\pi}\frac{c}{r_e  \omega}\right)^{3/5}\left(\frac{\omega_B}{\omega}\right)^{1/5}\sigma_T.
	\label{sigmaO}
\end{equation}
The scattering cross section in an X-mode wave can be given by
\begin{equation}
	\sigma_{X}\sim\gamma_{\text{rrl},X}^2 \sigma_T\approx\left(\frac{3H_X}{2\pi}\frac{c}{r_e a_0 \omega}\right)^{3/4}\left(\frac{\omega_B}{\omega}\right)^{1/2} \sigma_T.
	\label{sigmaX}
\end{equation}
Note that the scattering cross section in Eq. (\ref{sigmaO}) does not depend on the wave intensity, which is a unique property for the intense O-mode wave. One can naturally define the ratio of these two cross sections,
\begin{equation}
	f_\sigma=\frac{\sigma_O}{\sigma_X}\approx H_O^{3/5}H_X^{-3/4}\left(\frac{a_c}{a_0}\right)^{-3/4}\approx\left(\frac{a_c}{a_0}\right)^{-3/4}.
	\label{fsigma}
\end{equation}
When $a_0<a_c$, one has $f_\sigma< 1$. This implies that the scattering cross section of a charged particle in an O-mode wave is smaller than that in an X-mode wave.

The aforementioned calculations assume that the propagation direction is perpendicular to the background magnetic field. For a wave propagating obliquely, the scattering process can be considered in a frame that moves relativistically along the background magnetic field. In order to keep the propagation direction perpendicular to the background magnetic field in the moving frame, the Lorentz factor of this frame should be
\begin{equation}
	\gamma_f=\frac{1}{\sin{\theta}},
\end{equation}
where $\theta$ is the angle between the propagation direction and the background magnetic field in the lab frame. The wave frequency and cyclotron frequency in the moving frame are (a prime symbol denotes the quantity in the moving frame)
\begin{equation}
	\omega^\prime=\frac{\omega}{\gamma_f},\text{ }\omega_B^\prime=\omega_B.
\end{equation}
The condition $a_0^\prime<a_c^\prime$ therefore can be rewritten as
\begin{equation}
	a_0<a_c \gamma_f^{3/5},
\end{equation}
where $a_0=a_0^\prime$ is Lorentz invariant. The ratio between O-mode and X-mode scattering cross sections is then transformed into
\begin{equation}
	f_\sigma^\prime\approx (\sin{\theta})^{9/20}\left(\frac{a_c}{a_0}\right)^{-3/4}.
\end{equation}
One can see that in the moving frame, the ratio is further reduced, but does not show a significant difference with Eq. (\ref{fsigma}).

\section{Implications for FRBs}\label{ifrb}

\begin{figure}
	\begin{center}
		\subfigure{
			\includegraphics[width=0.38\textwidth]{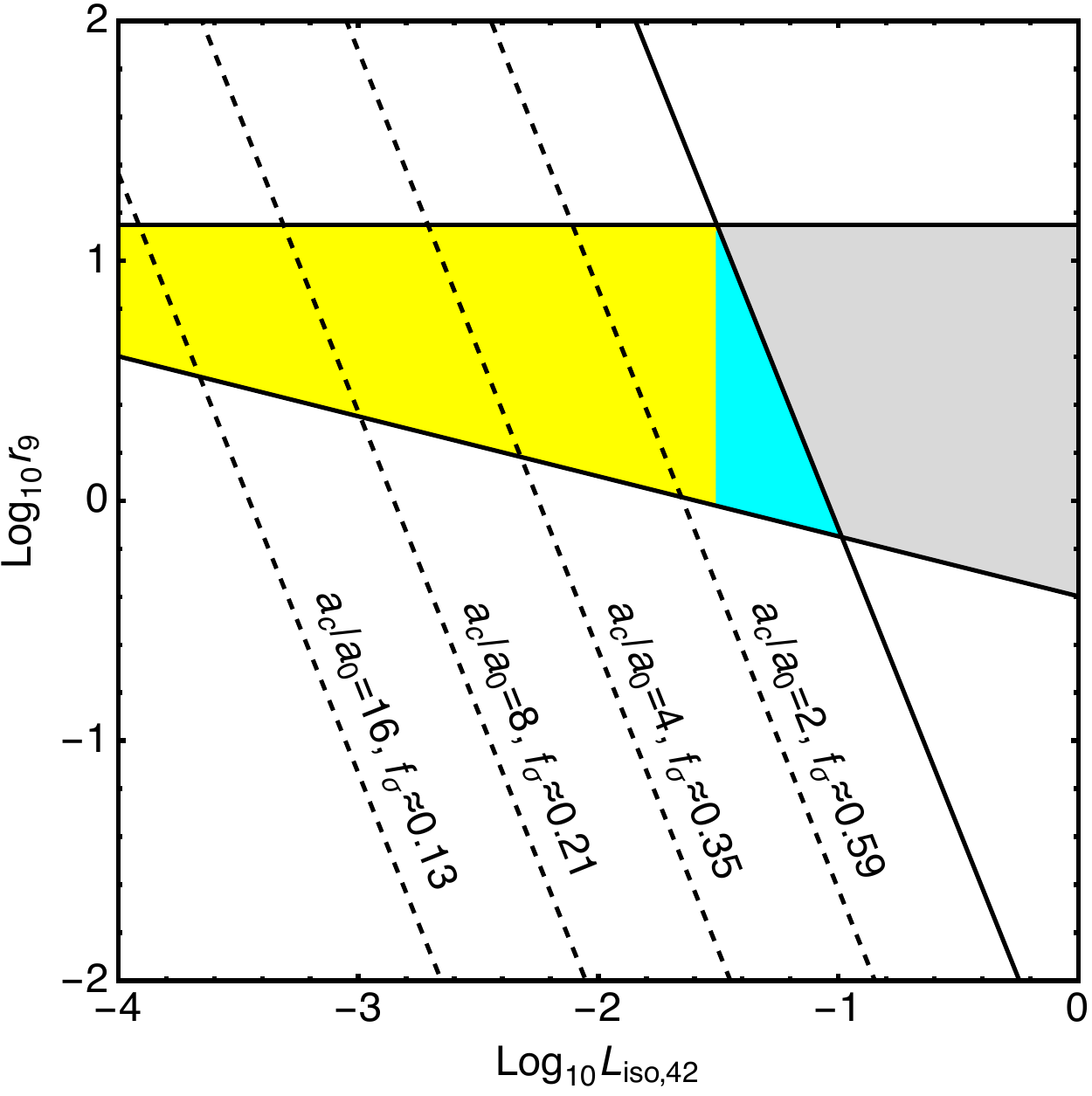}}
		\caption{The strong scattering region of FRBs is marked in yellow (part I) and cyan (part II) in luminosity--radius parameter space. Dashed lines represent various values of $a_c/a_0$ or $f_\sigma$. In the region marked in gray, radiation suppresses the growth of the Lorentz factor to the radiation reaction limit, and the scattering cross sections of particles in O-mode and X-mode waves are approximately equal. The parameters $B_{\text{s}}=10^{15}\text{ G}$ and $\nu=10^9\text{ Hz}$ are adopted.}	
		\label{L42r9}
	\end{center}
\end{figure}

If an FRB originates inside the magnetosphere, it is likely to be generated in the open magnetic field line region at a distance of several tens of neutron star radii. Therefore, the angle between the propagation direction of the FRB and the background magnetic field near the emission site is small \citep{Qu2022}. However, this angle is not extremely small near the outer magnetosphere. To simplify the calculation, we consider the case of perpendicular propagation. According to Eqs. (\ref{a0}) and (\ref{omegabomega}), the region where scattering is most severe, i.e., $1<\omega_B/\omega<a_0$ can be written as
\begin{equation}
	0.4 L_{\text{iso},42}^{-1/4}B_{\text{s},15}^{1/2}< r_9<14.1 \nu_{9}^{-1/3}B_{\text{s},15}^{1/3}.
	\label{detailregion}
\end{equation}
The condition $a_0<a_c$ further constrains the region where strong scattering occurs. For a typical FRB traveling through the magnetosphere of a magnetar, this can be written as
\begin{equation}
	\frac{a_c}{a_0}\approx0.3\nu_9^{2/5}L_{\text{iso},42}^{-1/2}B_{\text{s},15}^{2/5}r_9^{-1/5}>1.
	\label{aca0}
\end{equation}
Observations show that the luminosity of FRBs falls within a broad range of $10^{38}$--$10^{46}$ erg s$^{-1}$ \citep{Ravi2019,Bochenek2020}. Within such a wide range, the existence of a strong scattering region determined by both Eqs. (\ref{detailregion}) and (\ref{aca0}) is reasonable. We plot the region in the luminosity--radius parameter space in Figure \ref{L42r9}. Within this region, the strong scattering in different modes mentioned previously is valid. Different values of $a_c/a_0$ or $f_\sigma$ are marked by dashed lines in this figure. One can see that the scattering cross section of a particle in an O-mode wave is significantly smaller than that in an X-mode wave. In the gray region where $a_0>a_c$, the upper limit of the Lorentz factor from Eq. (\ref{gammaulimit}) is smaller than the Lorentz factor at the radiation reaction limit from Eqs. (\ref{gammarrlo}) and (\ref{gammarrlx}). Radiation then suppresses further growth of the Lorentz factor. The characteristic scattering cross sections of particles in O-mode and X-mode waves can be considered approximately equal since the maximal achievable Lorentz factors are equal in both modes.

\begin{figure}
	\begin{center}
		\subfigure{
			\includegraphics[width=0.38\textwidth]{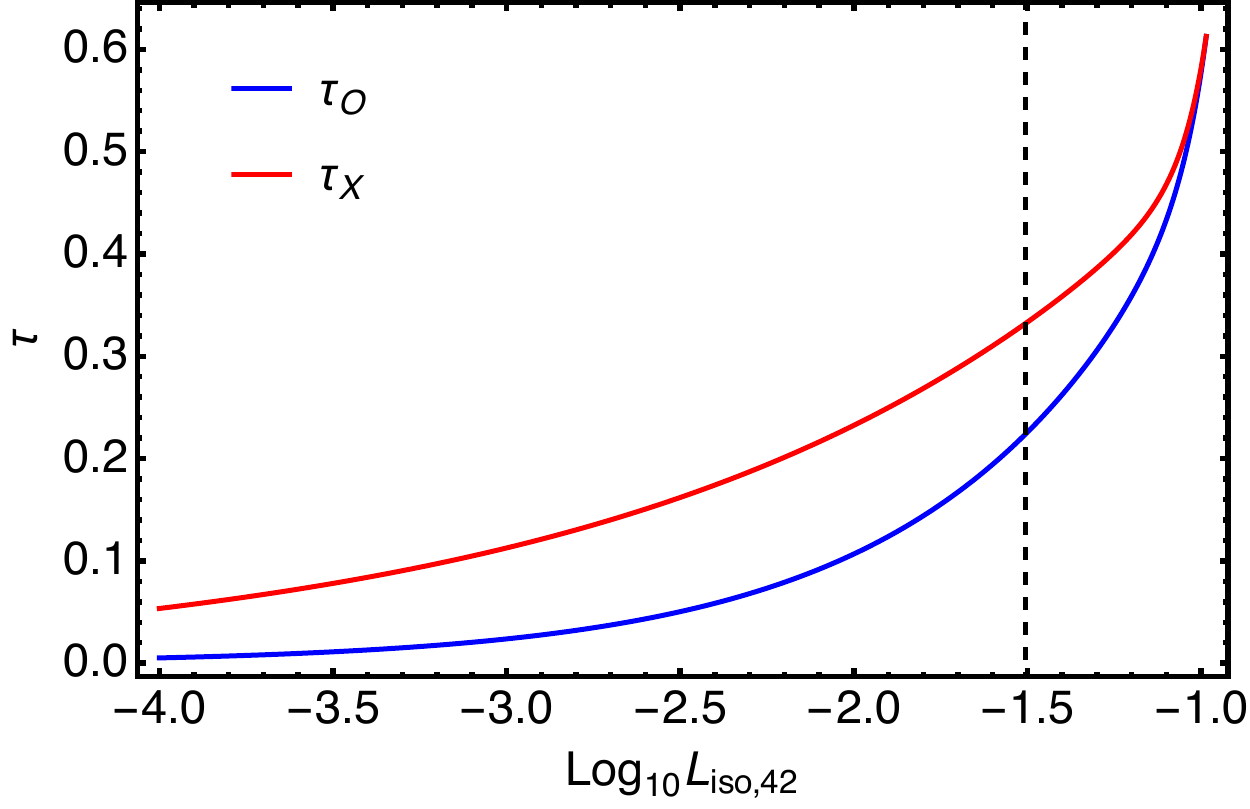}}
		\caption{Optical depth with respect to luminosity in the strong scattering region of FRBs. The black dashed line represents the boundary between part I and part II of the strong scattering region. The O-mode wave (blue) is more transparent than the X-mode wave (red). The parameters $\xi=10^2$, $B_\text{s}=10^{15}\text{ G}$,  $P=3 \text{ s}$, and $\nu=10^9\text{ Hz}$ are adopted.}	
		\label{tauL42}
	\end{center}
\end{figure}

The optical depth of an FRB in the strong scattering region is given by
\begin{equation}
	\tau_{O,\text{ }X}=\int \xi n_{\text{GJ}} \sigma_{O,\text{ }X} dr,
	\label{optdepth}
\end{equation}
where $n_{\text{GJ}}\approx7\times 10^4 \text{ cm}^{-3} \text{ }B_{\text{s},15} P^{-1} r_{9}^{-3}$ is the Goldreich-Julian density, and $\xi$ is the pair multiplicity parameter. In the following discussion, we fix the surface magnetic field strength $B_\text{s}=10^{15} \text{ G}$, the period $P=3\text{ s}$, and the wave frequency $\nu=10^9 \text{ Hz}$. The scattering cross section of a particle can then be written as $\sigma_O\approx3.3\times10^8 r_9^{-3/5}\sigma_T$ in an O-mode wave, and $\sigma_X\approx1.4\times10^8 L_{\text{iso},42}^{-3/8}r_9^{-3/4}\sigma_T$ in an X-mode wave. As shown in Figure {\ref{L42r9}}, since the upper bound of the strong scattering region is a piecewise function, we divide the region into part I and part II and calculate the value of optical depth for a given value of luminosity. In part I, the integral in Eq. (\ref{optdepth}) is calculated over $0.4L_{\text{iso},42}^{-1/4}<r_9<14.1$. In part II, the integral is calculated over $0.4L_{\text{iso},42}^{-1/4}<r_9<0.3^5 L_{\text{iso},42}^{-5/2}$. An FRB passing through part II will inevitably traverse the gray region, where the scattering cross section in both modes can be estimated as $\sigma_{O}^{\text{gray}}\approx\sigma_{X}^{\text{gray}}\approx\gamma_{\text{ul}}^2 \sigma_T\approx4.3\times10^{7}L_{\text{iso},42}^{-1}r_9^{-1}\sigma_T$. The integral in this region is calculated over $0.3^5 L_{\text{iso},42}^{-5/2}<r_9<14.1$. The optical depth as a function of luminosity is exhibited in Figure \ref{tauL42}. One can clearly see that due to different scattering cross sections, there is a significant difference between the values of optical depth of the O-mode and X-mode waves.

\section{Summary and Discussion}\label{discussion}

A novel semianalytical approach was developed to describe the behavior of a magnetized particle in a strong X-mode electromagnetic wave \citep{Beloborodov2022}. We have used a similar method and solved the motion of a magnetized particle in a strong O-mode wave. We found that the particle behaves similarly to that in the X-mode wave: it exchanges energy with the wave at frequent resonance events, and reaches the radiation reaction limit in a short time, leading to a significant enhancement in the scattering cross section. However, the pumped Lorentz factor of the particle at resonance in an O-mode wave exhibits a remarkable difference from that in an X-mode wave, as indicated by Eqs. (\ref{dgammaHO}) and (\ref{dgammaHx}). Consequently, the maximal Lorentz factors as well as scattering cross sections of particles in both modes are different. We further obtained the ratio of these two cross sections and found that this value is usually less than 1, implying that a charged particle in the X-mode wave is more likely to have a larger scattering cross section. We also briefly discussed the case of oblique propagation and found that it does not show a significant difference unless the angle between the magnetic field and the propagation direction is particularly small.

The above results have important implications if FRBs are generated inside the magnetosphere. When an FRB propagates outward, it likely encounters a strong scattering region determined by Eqs. (\ref{detailregion}) and (\ref{aca0}), as plotted in Figure \ref{L42r9}. The values of the optical depth of an O-mode wave and an X-mode wave are shown in Figure \ref{tauL42}. Our results indicate that an FRB is more transparent if it is an O-mode wave. We emphasize that this conclusion holds only in the strong scattering region, as it is possible for an FRB not to encounter such a region during its propagation.

The topic of whether an FRB can escape from the magnetosphere of a magnetar or not has been pointed out as an argument against the magnetospheric origin of FRBs. \cite{Beloborodov2021} argued that the photons emitted by charged particles in an FRB are capable of inducing the creation of electron--positron pairs, which increase the multiplicity parameter in the magnetosphere. This process then blocks the propagation of an FRB in the magnetosphere. However, this argument was challenged by several other perspectives. First, an FRB traveling in open field line regions probably reduces this effect \citep{Qu2022}. Second, the large radiation pressure possibly helps an FRB break out from the magnetosphere \citep{Wang2022}. Third, a self-cleaning effect may sweep away the plasma on the path of an FRB \citep{Lyutikov2023}. These effects can alleviate the criticism that an FRB can not escape the magnetosphere of a magnetar. In conclusion, we argue that if an FRB is partially scattered by magnetized particles, rather than being completely blocked,  an O-mode wave remains more transparent than an X-mode wave.

\section{Acknowledgments}
We are very grateful to the referee for the careful and thoughtful suggestions that have helped us improve this manuscript substantially. We thank Biao Zhang, Zi-Bin Zhang, Yuanhong Qu, Yue Wu, and Ze-Nan Liu for their useful discussions. This work is supported by  the National SKA Program of China (grant No. 2020SKA0120300) and National Natural Science Foundation of China (grant No. 12393812). S.Q.Z. acknowledges support from the National Natural Science Foundation of China (grant No. 12247144) and China Postdoctoral Science Foundation (grant Nos. 2021TQ0325 and 2022M723060).

\normalem
\bibliography{FRB}

\end{document}